\documentclass[
    ,final            
  ]
  {aipproc}

\usepackage{aas_macros,graphics}

\layoutstyle{6x9}

\newcommand{\beq}{\begin{equation}}
\newcommand{\eeq}{\end{equation}}

\begin{document}

\title{Energetics in MRI driven Turbulence}

\classification{95.30.Qd; 97.10.Gz}

\keywords{MHD; accretion disks}

\author{Thomas A. Gardiner}{
  address={Department of Astrophysical Sciences, 
           Princeton University, Princeton, NJ 08544}
}

\author{James M. Stone}{
  address={Department of Astrophysical Sciences,
           Princeton University, Princeton, NJ 08544}
}


\begin{abstract}
In these proceedings we present recent efforts to understand the
energetics of magnetohydrodynamic (MHD) turbulence driven by the
magneto-rotational instability (MRI).  These studies are carried out
in the local (shearing box) approximation using the Athena simulation
code.  Athena is a higher order Godunov algorithm based on the
piecewise parabolic method (PPM), the corner transport upwind (CTU)
integration algorithm, and the constrained transport (CT) algorithm
for evolving the magnetic field.  This algorithm is particularly
suited for these studies owing to the conservation properties of a
Godunov scheme and the particular implementation of the shearing box
source terms used here.  We present a variety of calculations which
may be compared directly to previously published results and discuss
them in some detail.  The only significant discrepancy found between
the results presented here and in the published literature involves
the turbulent heating rate.  We observe the presence of recurrent
channel solutions in calculations involving a mean vertical magnetic
field and the associated time lag between the energy injection and
thermalization rate.  We also present the results of a shearing box
calculation which includes an optically thin radiative term with a
cooling rate selected to match the turbulent heating rate.  Some
properties which we find uniformly present in all of the calculations
presented here are compressible fluctuations, spiral waves and weak
shocks.  It is found that these compressible modes dominate the
temporal fluctuations in the probability distribution functions for
most of the thermodynamic variables; only the specific entropy is
relatively immune to their effects.
\end{abstract}

\maketitle



\section{Introduction}

The last 10 years have borne witness to a substantial improvement in
our understanding of the role of sub-thermal magnetic fields in
accretion disk physics \cite{BH98RMP}.  The magneto-rotational
instability (MRI) plays the central role in driving
magnetohydrodynamic (MHD) turbulence which in turn transports mass and
angular momentum.  A large body of work now exists in the literature
which has elucidated a great many aspects of the MRI.  Yet
surprisingly, one aspect of accretion disk physics has yet to receive
significant attention; this is the role of energetics
\cite{Balbus2005ASP,BH2003LNP,Balbus2003}.

\par
One of the simplest energetics questions which could be addressed with
numerical simulations is, what energy reservoir is tapped, how is the
energy distributed amongst its kinetic and magnetic forms, and where
does it ultimately go.  Some elements of this question have been
answered repeatedly by a number of researchers.  The differential
rotation is the ultimate energy source which is tapped via the MRI and
injected into the kinetic and magnetic energy.  The energy stored in
the form of magnetic and kinetic energy may be exchanged
{}\cite{Brandenburg95}, but ultimately, it is dissipated and ends up
as thermal heating.  Interestingly enough, this last stage is
relatively unimportant as far as the MRI itself is concerned since
there is only a weak pressure dependence on the growth rates
{}\cite{Kim_Ostriker_2000} and saturation amplitudes \cite{Sano2004}.
Nevertheless, it is the last stage of this process which results in
radiation and allows a comparison between theoretical calculations and
observations.

\par
There are two circumstances under which the thermal state of the
plasma can be easily understood to have a significant impact on an MRI
turbulent accretion disk.  The first is in determining the vertical
structure of a stratified accretion disk.  The scale height of an
accretion disk is determined by balancing the thermal pressure
gradients against gravity.  The disk scale height is thus a measure of
the mean or central gas pressure, and is determined directly from the
balance of turbulent heating (as well as other potential sources) and
radiative losses.  (This is a question we intend to explore in the
near future.)  The second is by the strong temperature dependence on
the ionization state, resistivity, and other kinetic effects which
directly modify the induction equation such as ambipolar diffusion or
the Hall conductivity.  For example, the degree to which the MRI plays
a role in protostellar accretion will depend heavily on these factors
{}\cite{SGBH_PPIV,Kunz2004}.  It has also been suggested that the
temperature sensitivity of the resistivity could serve as a
\emph{switch} which regulates the outburst activity in dwarf nova disks
\cite{Gammie_Menou_98,Sano2003}.

\par
The focus of this paper is twofold with the ultimate goal to seek a
better understanding of the energetics in MRI turbulent accretion
disks.  This goal is ultimately achievable as a result of the recent
development of a fully second order accurate, conservative Godunov
algorithm for ideal MHD \cite{GS05}.  Using the shearing box
formalism, one is assured that any energy injected into the
computational domain ultimately results in thermally heating the
plasma without directly modeling the physical dissipation processes.
Nevertheless, since this is a new computational algorithm, and is the
first time that a fully second order Godunov algorithm has been
applied to the study of the MRI we are compelled to repeat a series of
calculations which can be directly compared with published results.
The second goal of this paper is to validate this new computational
algorithm when applied to the study of MRI driven MHD turbulence in
the shearing box and to understand any differences which arise between
the results presented here and published in the literature.  To that
end, we present results for adiabatic calculations with three
different field geometries and a calculation including radiative
cooling.


\section{Numerical Calculations}


\subsection{Shearing Box Approximation}

The studies presented here make use of the local ``shearing box''
formalism \cite{HGB95}.  In this approach, one focuses on a local
section of an accretion disk at some fiducial radius $R_0$ and in a
reference frame co-rotating at the local angular velocity $\Omega_0$.
Expanding the ideal MHD system of equations in cylindrical coordinates
for small deviations $\delta r/R_0 \ll 1$, etc. and relabeling the
coordinates $(dr, R_0 d\phi, dz)$ as $(dx, dy, dz)$ one obtains the
ideal MHD system in a Cartesian domain plus and additional set of
source terms which represent the Coriolis and tidal gravity forces.
For a Keplerian accretion disk one obtains
\begin{eqnarray}
\frac{\partial \rho}{\partial t} + 
{\bf\nabla\cdot} \left(\rho{\bf v}\right) & = & 0 
\label{eq:cons_mass} \\
\frac{\partial \rho {\bf v}}{\partial t} + 
{\bf\nabla\cdot} \left(\rho{\bf vv} - {\bf BB}\right) +
{\bf \nabla} P^* & = & 
\rho \Omega_0^2 (3 x \hat{i} - z \hat{k}) 
- 2 \Omega_0 \hat{k}\times \rho\bf{v} \\
\label{eq:cons_momentum}
\frac{\partial {\bf B}}{\partial t} + 
{\bf\nabla\cdot} \left({\bf v B - B v}\right) & = & 0 \\
\label{eq:cons_B}
\frac{\partial E}{\partial t} + 
\nabla\cdot((E + P^*) {\bf v} - {\bf B} ({\bf B \cdot v})) & = & 
\Omega_0^2 \rho{\bf v} \cdot (3 x \hat{i} - z \hat{k}) 
\label{eq:cons_energy}
\end{eqnarray}
where $\rho$ is the mass density, $\rho{\bf v}$ is the momentum
density, ${\bf B}$ is the magnetic field, and $E$ is the total energy
density.  The total pressure $P^* \equiv P + ({\bf B \cdot B})/2$ where $P$
is the gas pressure, and the total energy density $E$ is related to the
internal energy density $\epsilon$ via
\beq
E \equiv \epsilon + \rho({\bf v \cdot v})/2 + ({\bf B \cdot B})/2 ~.
\eeq
This system of equations is closed with an equation of state relating
the pressure to the density and internal energy.  We use an ideal gas
equation of state for which $P = (\gamma - 1) \epsilon$, and take
$\gamma=5/3$.  In the calculations presented in this paper we further
simplify this system of equations by neglecting the vertical
$z$-component of gravity.  As a result, the calculations presented
here are applicable to the central $\sim 1$ scale height of the
accretion disk.

\par
We present calculations both with and without radiative cooling.  In
those calculations which include optically thin radiative cooling we
choose a simple form for the cooling term which is appropriate for
bremsstrahlung radiation.  In these cases we add a term to the right
hand side of equation (\ref{eq:cons_energy}) of the form
\beq
-A \rho^2 (\epsilon/\rho)^{1/2}
\eeq
and choose the coefficient $A$ such that the cooling rate matches the
turbulent heating rate.


\subsection{Numerical Method}

The numerical method we use to solve the MHD equations in the shearing
box approximation can be described as having three basic parts.  These
are the basic integration algorithm, the treatment of source terms
within this algorithm and the application of the boundary conditions.
We briefly describe each of these in turn in what follows.


\subsubsection{Integration Algorithm}

We solve the shearing box system of MHD equations using the recently
developed simulation code, Athena.  This code was constructed to solve
the system of ideal hydrodynamics (HD) and magnetohydrodynamics (MHD)
by combining a few key algorithms.  The integration algorithm is based
upon the directionally unsplit corner transport upwind (CTU)
integration algorithm of Collela \cite{Collela_CTU,Saltzman_CTU}.  The
CTU integration algorithm can be formally understood as a procedure
for correcting the interface states in the piecewise parabolic method
(PPM) \cite{CW_PPM_84} to account for multidimensional wave
propagation.  We have combined this algorithm with the method of
constrained transport (CT) \cite{Evans_Hawley_CT} to evolve the
magnetic field and explicitly preserve the ${\bf\nabla\cdot B}=0$
condition.  A detailed description of the construction of this
integration algorithm for MHD in two dimensions is in press
\cite{GS05} and a description of the three dimensional algorithm is in
preparation.

\par
At the time of this writing, detailed code verification calculations,
user and programmer's guides, download information, etc. can be found
on the Athena home page at
{}\url{http://www.astro.princeton.edu/~jstone/athena/}.  See also the
article by J. Stone in these proceedings for more information on the
Athena integration algorithm and a description of the first
application results.


\subsubsection{Source Terms}

\par
One of our principal goals in this work is to study the energetics in
MRI turbulent accretion disks.  Hence, the evolution of the energy
equation (\ref{eq:cons_energy}) is of particular interest.  Energy
conservation can be restored to this system of equations by solving
for the evolution of the \emph{total} energy $E_t=(E+\rho\Phi)$,
including the tidal potential, $\Phi=-1.5\Omega_0^2x^2$.  With this
choice, one is guaranteed by the conservative nature of the equations,
that energy can be exchanged between its kinetic, magnetic, and
thermal forms, but cannot be lost in the form of truncation error.
The only route by which energy can be added to or removed from the
computational domain is through the boundaries.  In fact, making use
of the shearing periodic boundary conditions one can show \cite{HGB95}
that the volume average of the total energy $<E+\rho\Phi>$ evolves as
\beq
\frac{\partial}{\partial t} <E+\rho\Phi> = \frac{1.5 \Omega_0}{L_y L_z}
\int_X \left(\rho v_x \delta v_y - B_x B_y\right) dy dz
\label{eq:energy_evolution}
\eeq
where $\delta v_y = v_y + 1.5\Omega_0 x$ is the angular velocity
fluctuation, and the integral is taken over one of the bounding
$x$-faces.  This is precisely the route by which the MRI taps the
kinetic energy in the form of differential rotation in the shearing
box and amplifies the magnetic field.  In the calculations presented
here, the integral relation in equation (\ref{eq:energy_evolution}) is
satisfied to numerical roundoff error.

\par
We have found that the treatment of the source terms in the momentum
equation (\ref{eq:cons_momentum}) is particularly important for
shearing box calculations with Athena.  Note that unlike the energy
equation, the momentum equation cannot be recast in a conservative
form.  Moreover, there is a simple limit in which the $x$- and
$y$-momentum equations are \emph{strongly} coupled via the source
terms and therefore should not be evolved independently.  This strong
coupling limit describes the solution for uniform epicyclic
oscillation modes.  For epicyclic oscillations there is a conserved
kinetic energy
\beq
E_{epi} = \frac{1}{2} \rho \left( v_x^2 + 4 (\delta v_y)^2\right) ~.
\eeq
We have found that implementing the source terms in Athena in such a
way that the epicyclic kinetic energy is conserved is extremely
important.

\par
In order to clarify why the conservation of the epicyclic kinetic
energy should have such a large impact on the solution, note that it
is the long wavelength modes which contain a dominant share of the
energy.  This is a natural consequence of the shearing box formalism
in which the driving scale is effectively equal to the radial size of
the computational domain and the dissipation which occurs in the form
of a turbulent cascade.  In addition to long wavelength modes which
fit into the computational domain, the shearing box MHD equations (and
boundary conditions) also support uniform epicyclic oscillation modes.
It is reasonable to expect that some coupling between long wavelength
modes (which fit in the computational domain) and epicyclic
oscillation modes may result from numerical truncation error.  By
conserving the epicyclic kinetic energy, one ensures that this
coupling can not set up a positive feedback loop amplifying the
kinetic energy in the box.

\par
In order to conserve the epicyclic kinetic energy we evolve not the
$y$-momentum equation, but the equivalent conservation law for the
angular momentum fluctuation $\rho\delta v_y$.  When written in this
form, uniform epicyclic motion is consistent with the vanishing of the
flux gradient terms.  This leaves the $x$- and $y$-momentum
fluctuations evolution to be controlled by the resulting source terms.
One can show that evaluating the source terms using Crank-Nicholson
preserves the epicyclic kinetic energy.

\par
For those calculations which include radiative cooling, we apply the
operator splitting technique and split the radiative cooling equation
\beq
\frac{\partial \epsilon}{\partial t} = -A \rho^2 (\epsilon/\rho)^{1/2}
\eeq
from the rest of the shearing box evolution equations and solve for
the radiative cooling at constant density.

\subsubsection{Boundary Conditions}

The boundary conditions which we employ in these calculations are
strictly periodic in the $y$- and $z$-directions and shearing periodic
in the $x$-direction.  One result of these boundary conditions is that
so long as the volume averaged $x$-component of the magnetic field is
zero, the mean flux through the computational domain is fixed in time.
The implementation of the boundary conditions in this work preserves
this condition to roundoff error for the $x$- and $z$-components of
the magnetic field and to truncation error for the $y$-component of
the magnetic field.  Moreover, the remapping of the magnetic field in
the ghost zones at the $x$-boundaries of the computational domain
preserves the ${\bf \nabla\cdot B}=0$ condition to roundoff error.
The details of how this is achieved will be described in a future
paper.


\section{Results}

In this section we will present results from three local shearing box
simulations with different magnetic field geometries.  The cases which
we consider here include a mean toroidal field, a mean vertical field,
and a zero net-flux case.  Our choice of computational domain and
initial conditions mirrors that used in \cite{HGB95}.  Namely, we
choose a computational domain given by $-0.5\le x \le0.5$, $-\pi\le y
\le\pi$, $-0.5\le z \le0.5$ and resolve it on a $64\times128\times64$
grid, except where stated otherwise.  The initial velocity components
$v_x=v_z=0$ and $v_y=-1.5\Omega_0 x$.  We also choose
$\Omega_0=10^{-3}$, the initial density $\rho=1$ and pressure
$P=10^{-6}$.  To these initial conditions we add $0.1\%$ white noise
adiabatic density and pressure perturbations.  For reference, in the
following analysis we choose the zero point of the tidal potential
$\Phi=(0.125 - 1.5x^2)\Omega_0^2$ so that initially the mean tidal
potential energy $<\rho\Phi>=0$.


\subsection{Uniform Toroidal Magnetic Field}

\begin{figure}
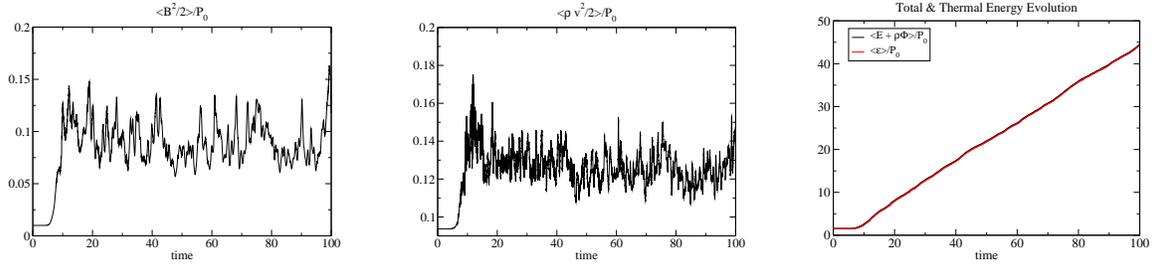

\includegraphics*[width=.3\textwidth]{Mag_Energy.y2.eps}
 \hspace{0.05\textwidth}
\includegraphics*[width=.3\textwidth]{Kin_Energy.y2.eps}
 \hspace{0.05\textwidth}
\includegraphics*[width=.3\textwidth]{Total_Therm_Energy.y2.eps}
\caption{Time evolution of the volume averaged magnetic, kinetic, total and 
thermal energy for the mean toroidal field calculation Y2.  Time is in
units of the orbital period.}
\label{fig:y2-energy}
\end{figure}
\begin{figure}
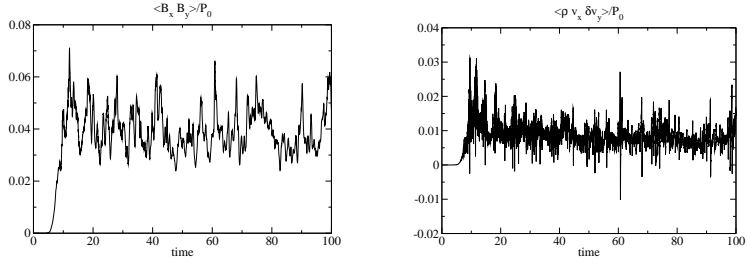

\includegraphics*[width=.3\textwidth]{Max_Stress.y2.eps}
 \hspace{0.05\textwidth}
\includegraphics*[width=.3\textwidth]{Reyn_Stress.y2.eps}
\caption{Time evolution of the volume averaged Maxwell and Reynolds
stress for the mean toroidal field calculation Y2.  Time is in units
of the orbital period.}
\label{fig:y2-stress}
\end{figure}

\par
The first case we consider has an initial magnetic field given by $B_x
= B_z = 0$ and $B_y = const.$ with $\beta = 2P/B^2=100$.  This model
calculation can be compared directly to models Y1 and Y11 in
\cite{HGB95}. In figure \ref{fig:y2-stress} we present the time
history of the volume averaged Maxwell and Reynolds stress.  The
volume averaged Maxwell stress $<-B_x B_y>/P_0$ and Reynolds stress
$<\rho v_x \delta v_y>/P_0$ have a mean value of $0.040 \pm 8.0\times
10^{-3}$ and $8.7\times 10^{-3} \pm 3.4\times 10^{-3}$ respectively
where the error is given by one standard deviation.  The ratio of the
Maxwell to Reynolds stress is $\sim 4.4$.  These values are in good
agreement with the results of \cite{HGB95}.  Similarly we find that
the saturation value for the kinetic and magnetic energy density shown
in figure \ref{fig:y2-energy} are in good agreement with the results
of \cite{HGB95}.

\par
The results presented here differ from those in \cite{HGB95} when we
consider the evolution of the thermal energy density.  In figure
\ref{fig:y2-energy} we present the evolution of the volume averaged
thermal energy density $<\epsilon>$ and the total energy density $<E +
\rho\Phi>$.  As noted previously, the total energy in the shearing box
is conserved and its volume average can increase only through the
differences of the flux across the radial boundaries.  From figure
\ref{fig:y2-energy} we see that the internal and total energies track
one another very closely as they must after the kinetic and magnetic
energy densities reach saturation.  Applying a linear curve fit to the
evolution of the total energy density after saturation and applying
equation (\ref{eq:energy_evolution}), one may extract the surface
averaged value of the Maxwell plus Reynolds stress one finds a very
good agreement with the volume averages noted above.  When compared to
the results in \cite{HGB95} we find that the turbulent heating rate is
increased by nearly an order of magnitude.  This difference is likely
attributable to the lack of a resistive heating term in the
evolutionary equation for the internal energy.


\subsection{Uniform Vertical Magnetic Field}

\begin{figure}
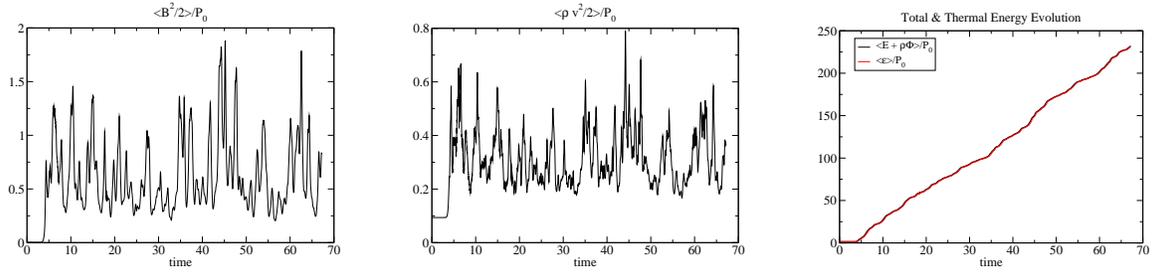

\includegraphics*[width=.3\textwidth]{Mag_Energy.z2.eps}
 \hspace{0.05\textwidth}
\includegraphics*[width=.3\textwidth]{Kin_Energy.z2.eps}
 \hspace{0.05\textwidth}
\includegraphics*[width=.3\textwidth]{Total_Therm_Energy.z2.eps}
\caption{Time evolution of the volume averaged magnetic, kinetic, total and 
thermal energy for the mean vertical field calculation Z2.  Time is in
units of the orbital period.}
\label{fig:z2-energy}
\end{figure}
\begin{figure}
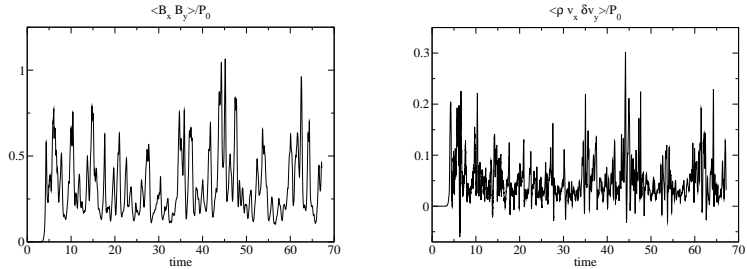

\includegraphics*[width=.3\textwidth]{Max_Stress.z2.eps}
 \hspace{0.05\textwidth}
\includegraphics*[width=.3\textwidth]{Reyn_Stress.z2.eps}
\caption{Time evolution of the volume averaged Maxwell and Reynolds
stress for the mean vertical field calculation Z2.  Time is in units
of the orbital period.}
\label{fig:z2-stress}
\end{figure}

\par
Next we consider the case where the magnetic field is initialized with
$B_x=B_y=0$ and $B_z=const.$ with $\beta=2P/B^2=400$.  This model
calculation can be compared directly to models Z4, Z7, Z19 and Z22 in
\cite{HGB95}.  In figure \ref{fig:z2-energy} we present the temporal
evolution of the volume averaged kinetic, magnetic, thermal and total
energies, and in figure \ref{fig:z2-stress} the Maxwell and Reynolds
stress.  It is immediately apparent upon inspection that this field
configuration is much more time variable as a result of recurrent
channel flows \cite{Sano2001}.  The volume averaged Maxwell stress
$<-B_x B_y>/P_0$ and Reynolds stress $<\rho v_x \delta v_y>/P_0$ have
a mean value of $0.33 \pm 0.18$ and $0.051 \pm 0.040$ respectively
where the error is given by one standard deviation.  These values are
approximately 4 times larger than for the mean toroidal field
configuration.  The ratio of the Maxwell to Reynolds stress is $~6.4$.
These numbers are in good agreement with \cite{HGB95}.

\par
The evolution of the volume averaged internal energy $<\epsilon>$ in
figure \ref{fig:z2-energy} shows approximately a factor of 7 times
larger heating rate than was found in \cite{HGB95}.  It is interesting
to note that the time variability of the Maxwell and Reynolds stress
in these flows is sufficiently large that it results in a clearly
discernible stair stepping on the evolution of the total and internal
energy density.

\par
\begin{figure}
\includegraphics*[width=.5\textwidth]{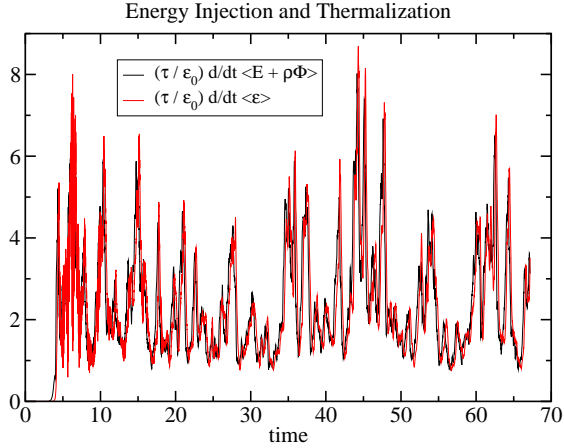}
\caption{Time evolution of the time derivative of the volume averaged 
total energy and thermal energy.  This figure demonstrates a clearly
discernible delay between the energy injection and thermalization
rate.  Time is in units of the orbital period.}
\label{fig:inject_thermalize}
\end{figure}

It was shown in \cite{Sano2001} that the presence of recurrent channel
flows results in a clearly observable time delay between the time
derivative of the volume average of the total energy $<E + \rho\Phi>$
and the time derivative of the volume average of the thermal energy
density $<\epsilon>$.  Note that these quantities are equivalent to
the energy injection rate and the energy thermalization rate.
Differentiating the total and thermal energy plots in figure
\ref{fig:z2-energy} and normalizing the derivative to the ratio of the
initial thermal energy density and the orbital period, one obtains an
equivalent plot, shown in figure \ref{fig:inject_thermalize}, to that
presented in \cite{Sano2001}.  Despite the presence of high frequency
oscillations in the thermal energy evolution (which are a result of
magnetosonic oscillations) there is a clearly discernible time lag
between the peak in the energy injection rate and the average
turbulent dissipation rate.  The results presented in figure
{}\ref{fig:inject_thermalize} also agree with \cite{Sano2001} in that
the thermalization rate overshoots the energy injection rate at its
maximum.  The fact that the results presented here are for an ideal
MHD fluid, while those in \cite{Sano2001} are for a magnetic
Reynolds number $\textrm{Re}_M=1$ provides strong support for both
calculations.  The time lag between the energy injection rate and
thermalization rate is a direct result of the channel flows and is
essentially independent of the nature of the resistivity.  Note that
this delay is not observed for any other field configuration.

%


\subsection{Zero Net Flux}

\par
\begin{figure}
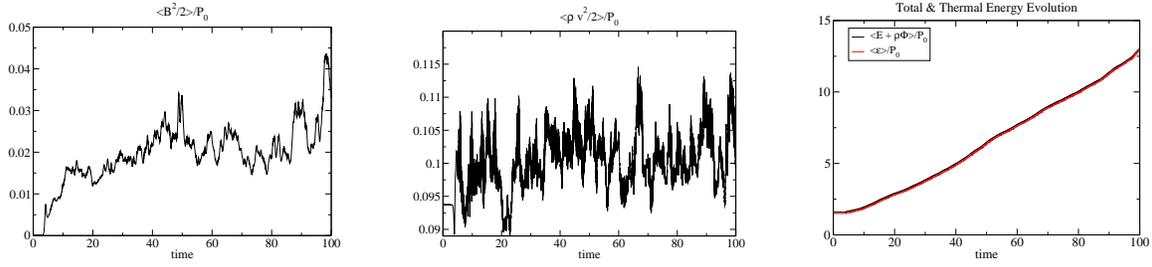

\includegraphics*[width=.3\textwidth]{Mag_Energy.s2.eps}
 \hspace{0.05\textwidth}
\includegraphics*[width=.3\textwidth]{Kin_Energy.s2.eps}
 \hspace{0.05\textwidth}
\includegraphics*[width=.3\textwidth]{Total_Therm_Energy.s2.eps}
\caption{Time evolution of the volume averaged magnetic, kinetic, total and
thermal energy for the zero net field calculation S2.  Time is in
units of the orbital period.}
\label{fig:s2-energy}
\end{figure}
\begin{figure}
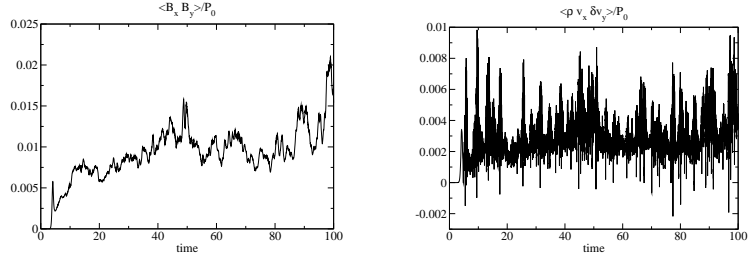

\includegraphics*[width=.3\textwidth]{Max_Stress.s2.eps}
 \hspace{0.05\textwidth}
\includegraphics*[width=.3\textwidth]{Reyn_Stress.s2.eps}
\caption{Time evolution of the volume averaged Maxwell and Reynolds
stress for the zero net field calculation S2.  Time is in units
of the orbital period.}
\label{fig:s2-stress}
\end{figure}

Next we consider the case where the magnetic field is initialized with
$B_x=B_y=0$ and $B_z=B_0 \sin(2\pi x/L_x)$ with $\beta=2P/B_0^2=4000$.
It was shown in \cite{HGB96} that the saturation of the MRI is
independent of the initial value of $\beta$.  As a result, this model
calculation can be compared to the results of models SZ1, SZ2 and SZ3
in \cite{HGB96}.  In figure \ref{fig:s2-energy} we present the
temporal evolution of the volume averaged kinetic, magnetic, thermal
and total energies, and in figure \ref{fig:s2-stress} the Maxwell and
Reynolds stress.  The volume averaged Maxwell stress $<-B_x B_y>/P_0$
and Reynolds stress $<\rho v_x \delta v_y>/P_0$ have a mean value of
$9.5 \times 10^{-3} \pm 2.8\times 10^{-3}$ and $2.9\times 10^{-3} \pm
1.3\times 10^{-3}$ respectively where the error is given by one
standard deviation. These values are approximately 4 times smaller
than for the mean toroidal field configuration.  The saturation
amplitude of the magnetic energy density agrees well with the results
of \cite{HGB96}.  Another feature of these flows is that they strongly
excite compressional fast magnetosonic waves which propagate both
inward and outward in the computational domain.  These compressional
waves are primarily responsible for the rapid time variability in the
mean Reynolds stress as shown in figure \ref{fig:s2-stress}.  The
presence of these waves was also noted in \cite{HGB96}.

%


\subsection{Compressibility Effects}

\begin{figure}
\includegraphics*[width=0.75\textwidth]{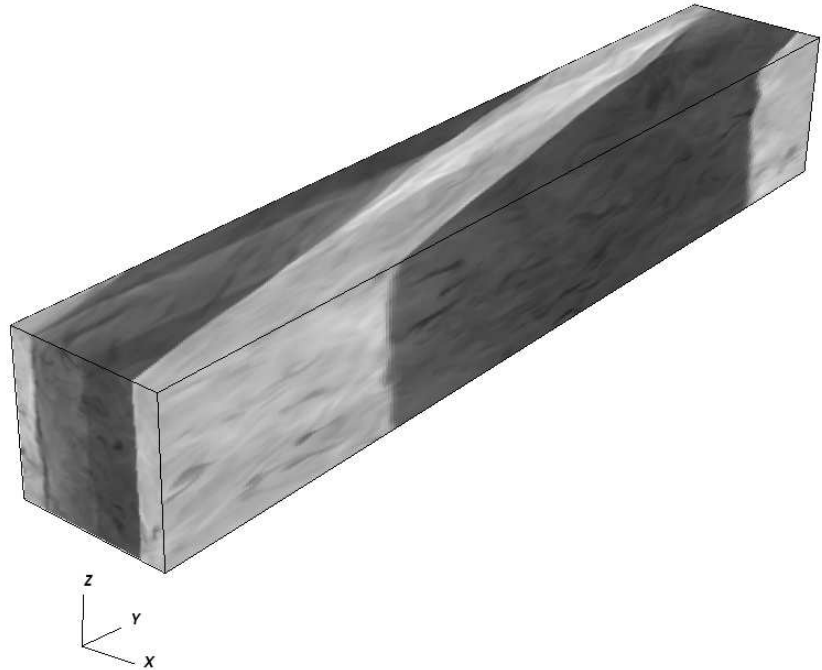}
\caption{Greyscale image of the density in model S0C1 showing the 
presence of spiral shock waves.  The density ranges from 0.8 (black)
to 1.2 (white).}
\label{fig:s0c1_density}
\end{figure}

\par
\begin{figure}
\includegraphics*[width=.5\textwidth]{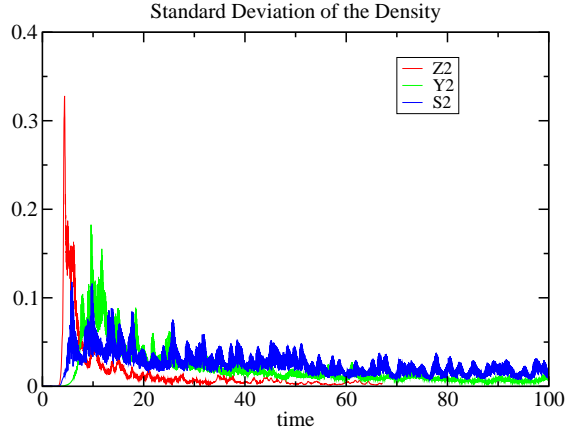}
\caption{Time evolution of the standard deviation of the density PDF 
for the three adiabatic calculations S2, Y2, and Z2.  Time is in units
of the orbital period.}
\label{fig:density_std}
\end{figure}

A rather dominant feature which we have observed in our shearing box
simulations is the formation of spiral waves and weak shocks.  Figure
\ref{fig:s0c1_density} shows an example of a weak shock in the density
in the model S0C1 calculation described in the following section.
Watching the time evolution of the density one finds the recurrent
formation of weak spiral shocks which rapidly propagate across the
grid, dissipate and at some later time are regenerated.  One way to
study this phenomena is to look at the probability distribution
function (PDF) for the density.  Perhaps the most illuminating aspect
of the density PDF is the time evolution of the standard deviation as
shown in figure \ref{fig:density_std}.  The first feature to notice in
this figure is that all three models show a rapid increase in the
density dispersion as the system makes the transition to a turbulent
state followed by a gradual decrease toward some small positive value.
Note that the larger the heating rate, the faster the standard
deviation of the density decreases.  These results support the
conclusion that the decrease in the density dispersion as the system
evolves is a simple consequence of the increasing thermal energy
density driving the system toward the incompressible limit.
Calculations initiated with a larger initial pressure also support
this conclusion.  Interspersed on this plot are regions of rapid
variability, which are unresolved in the figure, but have a well
defined characteristic period described by fast magnetosonic
oscillations with an oscillation period $\tau \approx L_x/c_{fast}$.
These coherent magnetosonic oscillations appear to be the direct
result of the localized dissipation of magnetic energy in current
sheets via ``numerical resistivity'' and the shearing box boundary
conditions which effectively make the shearing box a resonant cavity
like structure at wavelengths comparable to the box size.  In the Y2
and Z2 models at late times we find that the PDF for the density,
temperature, and specific entropy are well described by a Gaussian
distribution.  At early times, the density and temperature PDFs for
all calculations presented here are highly time variable resulting
from the compressible modes in the shearing box.

\par
It is worth pointing out that the initial pressure in the models
presented here was selected to correspond to approximately the central
1 scale height of a vertically stratified accretion disk.  In a time
steady accretion disk calculation, losses due to radiative cooling
will balance the turbulent heating and the compressibility will be
non-negligible.  These results suggest that the localized dissipation
of magnetic energy in current sheets and the resulting compressional
waves may play an important role determining the temperature
fluctuations and hence the spectra of accretion disks.

%


\subsection{Radiative Zero Net Flux}

\begin{figure}
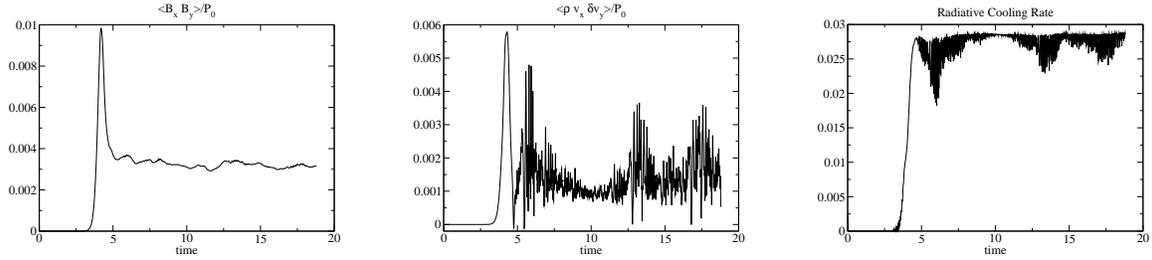

\includegraphics*[width=.3\textwidth]{Max_Stress.s0c1.eps}
 \hspace{0.05\textwidth}
\includegraphics*[width=.3\textwidth]{Reyn_Stress.s0c1.eps}
 \hspace{0.05\textwidth}
\includegraphics*[width=.3\textwidth]{Rad_Cool_Rate.s0c1.eps}
\caption{Time evolution of the volume averaged Maxwell and Reynolds
stress and the radiative cooling rate for the zero net field
calculation S0C1.  Time is in units of the orbital period and the
radiative cooling rate is in units of $(\epsilon_0/\tau)$, the initial
internal energy divided by the orbital period.}
\label{fig:s0c1-stress}
\end{figure}

In this section we present the results of a zero net flux calculation
including radiative cooling.  This calculation is identical to the
zero net flux calculation presented earlier with the exception that it
is resolved on a $128\times256\times128$ grid.  In figure
\ref{fig:s0c1-energy} we present the time evolution of the volume
averaged kinetic, magnetic, thermal and total energies.  The radiative
cooling rate was selected in this calculation to match the turbulent
heating rate in the saturated state.  As a result the total and
thermal energy density show little time evolution.  Note also the
highly time variable kinetic energy density which is a result of the
magnetosonic oscillations and spiral shock waves which are present in
this calculation.  In figure \ref{fig:s0c1-stress} we present the
volume average of the Maxwell and Reynolds stress.  The volume
averaged Maxwell stress $<-B_x B_y>/P_0$ and Reynolds stress $<\rho
v_x \delta v_y>/P_0$ have a mean value of $3.2 \times 10^{-3} \pm 1.7
\times 10^{-4}$ and $1.4 \times 10^{-3} \pm 6.4 \times 10^{-4}$
respectively where the error is given by one standard deviation.
These values are approximately a half of the values found in the
calculations presented earlier at half the resolution.  This strong
resolution dependence is not found in either the mean toroidal or mean
vertical field calculations.  Note that these values for the Maxwell
and Reynolds stress are not a result of the presence of radiative
cooling in these calculations, as we have also performed non-radiative
calculations at this resolution and find essentially identical values.

\par
In figure \ref{fig:s0c1-stress} we present the volume averaged
radiative cooling rate in units of $(\epsilon_0/\tau)$, the ratio of
the initial internal energy density to the orbital period.  Note that,
as a consistency check, one can relate the radiative cooling rate the
the mean Maxwell and Reynolds stress shown in figure
{}\ref{fig:s0c1-stress} via equation (\ref{eq:energy_evolution}) under
the assumption that the turbulent heating is balanced by the radiative
cooling.  As should be expected, one finds a very good agreement from
this test.  There are two main features present in this plot, the
rapid rise as the system makes the transition to the turbulent state,
and the oscillations resulting from the compressional magnetosonic
waves and spiral shocks.  While the high frequency oscillations are
strongly determined by the boundary conditions of the shearing box, it
is suggestive that the coupling of compressional modes to radiative
losses may have an impact on the observed spectrum from accretion
disks.

\begin{figure}
\includegraphics*[width=.3\textwidth]{Mag_Energy.s0c1.eps}
 \hspace{0.05\textwidth}
\includegraphics*[width=.3\textwidth]{Kin_Energy.s0c1.eps}
 \hspace{0.05\textwidth}
\includegraphics*[width=.3\textwidth]{Total_Therm_Energy.s0c1.eps}
\caption{Time evolution of the volume averaged magnetic, kinetic, total and
thermal energy for the zero net field calculation S0C1.  Time is in
units of the orbital period.}
\label{fig:s0c1-energy}
\end{figure}
%


\section{Conclusions}

We have performed a series of shearing box calculations for three
different field geometries with the recently developed simulation code
Athena.  This is the first time that a fully second order accurate
Godunov algorithm has been applied to the study of the
magnetorotational instability in the shearing box formalism.  As such,
comparing the calculations presented here to previously published
results provides an independent confirmation.  In general we have
found that for non-radiative calculations the saturation amplitudes of
the magnetic and kinetic energy, the Maxwell and Reynolds stress are
in good agreement with the results published in \cite{HGB95,HGB96}.
We also found from the non-radiative calculations that the energy
dissipation rate is underestimated in \cite{HGB95} and presumably also
in \cite{HGB96} by a factor of $7-10$.  This difference is likely
attributable to the lack of a resistive heating term in the
evolutionary equation for the internal energy.

\par
One of the consequences of energy conservation is the accurate
treatment of the dissipation of magnetic energy into thermal.  This
has been found to influence the solution in the shearing box in two
ways.  The first is the secular heating of the plasma as must occur
since over a long time average the energy injection and thermalization
rates must balance.  This has the tendency to push the system toward
an incompressible configuration in which the PDF of the thermodynamic
variable tends toward a Gaussian distribution.  The second is the
generation of compressible magnetosonic waves and spiral shocks.  This
is presumably a result of the fact that in ideal MHD calculations the
numerical resistivity is nonlinear leading to the dissipation of the
magnetic field predominantly in the form of current sheets.  This
localized heating process generates compressional waves which
effectively resonate in the shearing box. These compressional modes,
which are most important at early times, have a strong influence on
the PDF for the density and temperature in these simulations.  In
radiatively cooling calculations they also have a direct influence on
the radiative cooling rate.  These results are suggestive that
magnetic field dissipation in current sheets and the resulting
compressional modes may have an observable influence on an accretion
disk spectrum.

\par
We have also observed the recurrent channel solutions in simulations
with a mean vertical magnetic field and the associated time lag
between the energy injection rate and energy thermalization rate
described in \cite{Sano2001}.  The fact that the calculations
presented here were performed with ideal MHD, while those in
{}\cite{Sano2001} have a magnetic Reynolds number $\textrm{Re}_M=1$
provides strong support that these results are essentially independent
of the magnetic resistivity, so long as the MRI can operate.

\par
The energetics of turbulent accretion disks remains one of the least
well explored and most fruitful areas for future research in accretion
physics.  Not only does an accurate treatment of the energetics allow
for contact between theoretical models and observations, but it also
allows us to study the way in which the highly nonlinear temperature
dependence of the resistivity, Hall and ambipolar diffusion terms
influence the the nonlinear behavior of the MRI.  These results hold
potentially significant promise for improving our understanding of the
accretion process in protostellar accretion disks and outburst
phenomena in dwarf nova disks.  Even the modest question of how the
interplay between radiative cooling and turbulent heating influence
the vertical structure of an accretion disk remains an open question.



\begin{theacknowledgments}
The calculations presented here were performed on the beowolf cluster
in the Princeton dept. of Astrophysical Sciences.  We would like to
acknowledge the support of the NSF grant AST-0216105.
\end{theacknowledgments}



\bibliographystyle{aipproc}   

\bibliography{gardiner_mri,gardiner_numerics}

\IfFileExists{\jobname.bbl}{}
 {\typeout{}
  \typeout{******************************************}
  \typeout{** Please run "bibtex \jobname" to obtain}
  \typeout{** the bibliography and then re-run LaTeX}
  \typeout{** twice to fix the references!}
  \typeout{******************************************}
  \typeout{}
 }

\end{document}